\documentclass[conference]{IEEEtran}

\usepackage[T1]{fontenc}
\usepackage[utf8]{inputenc}
\usepackage{bm}
\usepackage[caption=false]{subfig}
\usepackage{tikz}
\usepackage{pgfplots}
\usepackage{amsmath}
\usepackage{dsfont}
\usepackage{amssymb}
\usepackage{graphicx}
\usepackage{cite}
\usepackage{fst}
\usepackage{tumcolor}
\usepackage{siunitx}
\usepackage{algorithm}
\usepackage{algpseudocode}
\usepackage{booktabs}
\usepackage{textcomp}
\usepackage[normalem]{ulem}
\pgfplotsset{compat=newest}

\algnewcommand\algorithmicinput{\textbf{INPUT:}}
\algnewcommand\INPUT{\item[\algorithmicinput]}

\newcommand{\SNR}{\text{SNR}}
\newcommand{\BMD}{\text{BMD}}

\DeclareSIUnit\bpcu{bpcu}

\usetikzlibrary{external,intersections,calc}
\newcommand\undermat[2]{%
  \makebox[0pt][l]{$\smash{\underbrace{\phantom{%
    \begin{matrix}#2\end{matrix}}}_{\text{$#1$}}}$}#2}

\IEEEoverridecommandlockouts    
    
\begin{document}

\title{Design of Robust, Protograph Based LDPC Codes for Rate-Adaptation via Probabilistic Shaping}

\author{\IEEEauthorblockN{Fabian Steiner\IEEEauthorrefmark{1}, Patrick Schulte\IEEEauthorrefmark{2}}\IEEEauthorblockA{\IEEEauthorrefmark{1}Fachgebiet Methoden der Signalverarbeitung, \IEEEauthorrefmark{2}Institute for Communications Engineering\\Technical University of Munich\\Email: \{fabian.steiner, patrick.schulte\}@tum.de}\thanks{F. Steiner was supported by the TUM--Institute for Advanced Study, funded by the German Excellence Initiative and the European Union Seventh Framework Program under grant agreement n\textdegree{ }291763. P. Schulte was supported by the German Federal Ministry of Education and Research in the framework of an Alexander von Humboldt Professorship.}}

\markboth{}{}%

\maketitle

\begin{abstract}
In this work, the design of robust, protograph-based low-density parity-check (LDPC) codes
for rate-adaptive communication via probabilistic shaping is considered. Recently,
probabilistic amplitude shaping (PAS) by Böcherer \emph{et al.} has been introduced
for capacity approaching and rate-adaptive communication with a bitwise-demapper
and binary decoder. Previous work by the authors considered the optimization of protograph based
LDPC codes for PAS and specific spectral efficiencies (SEs) to jointly 
optimize the LDPC code node degrees and the mapping of the coded bits to the bit-interleaved coded
modulation (BICM) bit-channels. We show that these codes tend to perform
poor when operated at other rates and propose the design of robust LDPC codes by
employing a min-max approach in the search for good protograph ensembles via
differential evolution. The considered design uses a single 16 amplitude-shift-keying (ASK)
constellation and a robust $\num{13/16}\approx\num{0.813}$ rate LDPC code to operate between
\num{0.7} to \num{2.7} bits per channel use. For a blocklength of \num{16224} bits and
a target frame error rate of \num{e-3} the proposed code operates within 
\SI{1.32}{dB} of continuous AWGN capacity for \SIrange{0.7}{1.3}{\bpcu} and within \SI{1.05}{dB}
for \SIrange{1.3}{2.7}{\bpcu}.
\end{abstract}

\section{Introduction}

The practical operation of communication systems requires to adapt the 
\ac{SE} to the channel quality. For instance in optical systems, 
a transceiver that operates on a short network segment with high signal-to-noise 
ratio (SNR) should achieve a high spectral efficiency to maximize the net 
data rate over this segment. Similarly, a transceiver operating on a long network 
segment (e.g., an intercontinental route) with low SNR should use either 
a lower order modulation format or a \ac{FEC} code with low code rate to 
ensure reliable transmission. For wireless systems, rate adaptation is important,
as the channel quality often changes rapidly because of the user's mobility or changing 
fading conditions.

Current transceivers implement rate adaptation by supporting several \emph{modcods}, 
i.e., combinations of modulation formats and coding rates. For instance, 
LTE chooses from a set of 29 different modcods~\cite[Table~7.1.7.1-1]{etsi2013lte} and DVB-S2X~\cite{etsi2014dvb} 
defines 116 modcods~\cite[Table~1]{etsi2014dvb} extending the 40 modcods in DVB-S2~\cite{etsi2009dvb}.
Here, flexibility comes at the price of increased complexity and implementation 
overhead.

In~\cite{bocherer_bandwidth_2015}, seamless rate adaptation from \numrange{2}{10} bits per channel use (\si{\bpcu})
was demonstrated with only five modcods. This was achieved by \ac{PAS}, i.e., the concatenation
of a flexible distribution matcher~\cite{schulte_constant_2016} with a fixed off-the-shelf
DVB-S2 LDPC code. Its practical applicability was shown in optical experiments in~\cite{buchali_experimental_2015,buchali_rate_2016}.

As the coded performance is heavily influenced by the mapping of coded bits to the variable nodes of the
\ac{LDPC} code, using off-the-shelf codes usually requires to optimize the bit-mapper (see \cite[Sec. VIII]{bocherer_bandwidth_2015}
and references therein). To avoid this and further improve the performance, we designed 
protograph-based LDPC codes in~\cite{steiner_protographbased_2016}, which take by construction 
the bit-mapping into account. For an \ac{SE} of \SI{4.25}{\bpcu}, 
the obtained codes operate within \SI{0.69}{dB} of \ac{AWGN} channel capacity at a \ac{FER}
of \num{e-3} with a blocklength of \num{64800} bits.
As the codes in~\cite{steiner_protographbased_2016} are highly optimized for one particular \ac{SE},
their applicability in rate adaptive transceivers is limited. In the present work, we consider
robust \ac{LDPC} code design for rate adaptive transceivers. We design a rate 13/16 protograph-based~\cite{thorpe_protograph} 
LDPC code that can be used on a 16-ASK constellation to operate over the \ac{AWGN} channel with any \ac{SE} 
in the range from \SIrange{0.7}{2.7}{\bpcu}. We use the surrogate channel approach of~\cite{steiner_protographbased_2016} and develop
a new optimization metric for the differential evolution to find good protograph ensembles. For a target 
\ac{FER} of \num{e-3} and blocklength of \num{16224} bits, the considered system is able to operate within 
\SI{1.32}{dB} of continuous AWGN capacity for \SIrange{0.7}{1.3}{\bpcu} and within \SI{1.05}{dB}
for \SIrange{1.3}{2.7}{\bpcu}. The gap to Gallager's random coding bound decreases from \SI{0.68}{dB} for
low \acp{SE} to \SI{0.51}{dB} for larger \acp{SE}.

This work is organized as follows: Sec.~\ref{sec:system_model} reviews 
the transceiver model and we develop the \ac{LDPC} code design 
in Sec.~\ref{sec:ldpc_code_design}. Simulation results and a comparison to an off-the-shelf 
code are shown in Sec.~\ref{sec:sim_results}. We conclude in Sec.~\ref{sec:conclusion}.



\begin{figure*}
\footnotesize
\centering
\includegraphics{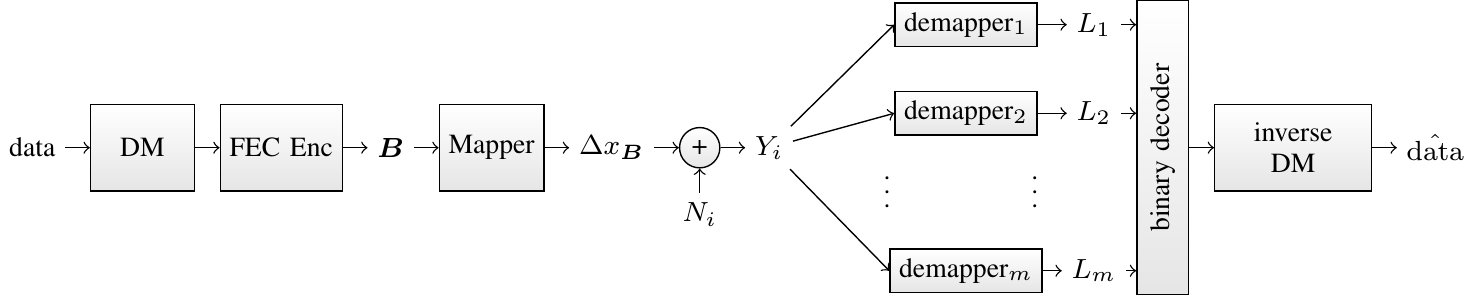}
\caption{Resulting channel model of the bit-metric decoding receiver structure.}
\label{fig:bmd_channel_model}
\end{figure*}

\section{System Model~\cite{bocherer_bandwidth_2015}}
\label{sec:system_model}

We consider transmission over a time discrete \ac{AWGN} channel
\begin{equation}
    Y_i = \Delta X_i + N_i\label{eq:awgn_channel}
\end{equation}
where the channel input $X_i$ has distribution $P_X$ on the $2^m$-\ac{ASK} constellation 
$\cX = \left\{\pm1,\pm3,\ldots,\pm2^m-1\right\}$ and the noise $N_i \sim \mathcal{N}(0,1)$ is 
Gaussian with zero mean and unit variance. We use the constellation scaling $\Delta \in\setR$ to 
set the $\SNR=\E{(\Delta X_i)^2}/1$.
Using a uniform distribution on \ac{ASK} constellations results in a shaping gap of \SI{1.53}{dB} in 
the high SNR regime~\cite[Sec.~IV-B]{forney_efficient_1984} compared to the capacity of the \ac{AWGN} channel
\begin{equation}
    C_\tawgn(\SNR) = \frac{1}{2}\log_2(1+\SNR)\label{eq:awgn_capacity},
\end{equation}
which is achieved with Gaussian inputs of zero mean and variance $\SNR$. This loss can be overcome 
by employing probabilistic shaping, i.e., imposing a non-uniform distribution on the \ac{ASK} constellation 
points~\cite[Sec.~III-D]{bocherer_bandwidth_2015}.

\subsection{Transmitter: Probabilistic Amplitude Shaping}
\label{sec:pas}

In order to combine probabilistic shaping with \ac{FEC}, we employ the \ac{PAS} scheme~\cite[Sec.~IV]{bocherer_bandwidth_2015}.
The optimal distribution $P_X^*$ for the AWGN channel is symmetric around the origin, 
i.e.,  $P_X^*(x) = P_X^*(-x)$. This allows for factorization into amplitude and sign, i.e.,
\[
P_X(x) = P_A(\abs{x})P_S(\sign(x)),
\]
where the amplitude $A$ is defined on $\cA = \set{1,3,\ldots,2^m-1}$ 
and $S=\sign(X)$ is uniformly distributed on $\{-1,+1\}$. For the binary labeling of the constellation points, we use
\[
\text{label}(x) = \beta(\sign(x))\vbeta(\abs{x})
\]
Herein, the function $\vbeta : \cA \to \set{0,1}^{m-1}$ implements a \ac{BRGC}~\cite{gray1953pulse} and $\beta(-1) = 0, \beta(+1) = 1$. 
Assuming a desired \ac{FEC} blocklength of $n$, a \ac{DM}~\cite{schulte_constant_2016} is used to transform 
uniformly distributed information bits into a sequence of $n_\tc=n/m$ amplitudes that follow a prescribed distribution and 
have entropy $\entr(A)$. A systematic encoder copies their binary representation $\vbeta(A^{n_\tc})$ together with potentially $(1-(1-c)\cdot m)\cdot n_\tc$ 
other uniformly distributed information bits to the final codeword and appends uniformly distributed check bits. Consequently,
we obtain an overall \ac{SE} $R$~\cite[Sec.~IV-D]{bocherer_bandwidth_2015} of
\begin{equation}
R = \entr(A) + 1-(1-c)\cdot m = \entr(X) - (1-c)\cdot m.\label{eq:transmission_rate}
\end{equation}
By changing the distribution $P_X$, the \ac{SE} $R$ can be adjusted seamlessly 
keeping the same modcod, i.e., the modulation order $2^m$ and the code rate $c$.
\begin{rem}
In case of uniform inputs, we have $\entr(X) = m$, such that \eqref{eq:transmission_rate} 
reduces to the well known \ac{SE} $c\cdot m$ of a system with uniform inputs.
\end{rem}

\subsection{Receiver: Bit-Metric Decoding (BMD)}
\label{sec:bmd}
\acused{BMD}

At the receiver side, we use \ac{BMD}, i.e., a bitwise demapper followed by a binary decoder.
This approach was introduced in \cite{zehavi_8-psk_1992} and is now mainly known as
\ac{BICM} \cite{caire_bit-interleaved_1998}. We identify each constellation symbol by 
its $m$ bit binary label $\vB=B_1B_2\ldots B_m$, where
\begin{align*}
 B_1 &= \beta(\sign(x)),\\
 B_2B_3\ldots B_m &= \vbeta(\abs{x}), \quad x\in\cX.
\end{align*}
Bit-metric decoding can be interpreted from a mismatched decoding perspective~\cite{martinez_mismatched}, 
where the decoder assumes both an auxiliary channel metric $q_{y|\vB} = \prod_{i=1}^m p_{Y|B_i}(y|b_i)$ 
and an auxiliary input distribution $Q_{\vB}(\vb)=\prod_{i=1}^m P_{B_i}(b_i)$. These assumptions represent a 
setting with $m$ parallel and independent channels of the form
\[
 p_{Y|B_i}(y|b) = \frac{1}{P_{B_i}(b)}\sum_{\xi\in\cX^b_i} p_{Y|X}(y|\xi)P_X(\xi),
\]
where $\cX^b_i$ denotes the set of symbols $x\in\cX$ with $B_i = b$.
The random variable $B_i$ has the marginal distribution
\[
 P_{B_i}(b) = \sum_{\xi\in\cX^b_i} P_X(\xi).
\]
$P_{B_i}(b)$ is both a function of the distribution $P_X$ and the labeling.
For each bit-level $i = 1,\ldots, m$ the soft-demapper then calculates its metric according to
\begin{equation}
 L_i = \log\left(\frac{P_{B_i}(0)}{P_{B_i}(1)}\right) + \log\left(\frac{p_{Y|B_i}(y|0)}{p_{Y|B_i}(y|1)}\right),
\end{equation}
which motivates our BICM channel model depicted in Fig.~\ref{fig:bmd_channel_model}.

In \cite{bocherer2014achievable}, a \ac{BMD}-achievable rate is given by
\begin{equation}
    R_\tbmd = \left[\entr(\vB) - \sum_{i=1}^m\entr(B_i|Y)\right]^+,\label{eq:bmd_rate}
\end{equation}
where $\left[\cdot\right]^+ = \max (0, \cdot)$.

We introduce the notation $R_\tbmd(P_X,\SNR)$ to denote the \ac{BMD} 
achievable rate with input distribution $P_X$ and an average power constraint of $\SNR$. We also introducde $R^{-1}_\tbmd(P_X,R)$ 
which describes the required $\SNR$ such that $R_\tbmd(P_X,\SNR)$ equals $R$.

\section{Robust LDPC Code Design}
\label{sec:ldpc_code_design}

We target the design of protograph based \ac{LDPC} codes that perform well over the range of \acp{SE} from \SIrange{0.7}{2.7}{\bpcu}. We will first show that
codes optimized for one specific \ac{SE} tend to perform bad at other rates. This behavior can be avoided by accounting for the whole range
during the code design. 

\subsection{Operating Points and Signaling}
We choose our operating points for different \acp{SE} by changing the entropy of the distribution 
on the constellation symbols. Following~\eqref{eq:transmission_rate} and considering a $c=13/16$ code and a
16-ASK constellation, the entropy and \ac{SE} are
\begin{equation}
\entr(X) = R + \left(1-\frac{13}{16}\right)\cdot 4 \quad\Longrightarrow\quad R = \entr(X) - \frac{3}{4}.\label{eq:entropy_constraint}
\end{equation}
Let $P_X^R$ be the \ac{MB} distribution~\cite{kschischang_pasupathy_maxwell} that leads to a \ac{SE} $R$ with a 13/16 rate code.

\begin{rem}
By~\eqref{eq:entropy_constraint} a rate constraint translates into an entropy constraint. The \ac{MB} distribution is the solution to the following optimization problem:
\begin{align*}
    \min_{\text{\ac{PMF}} P_X} \E{X^2} \quad \text{subject to }\quad \entr(X) \geq H'.
\end{align*}
Hence, it is a natural choice for power-efficient signaling with an entropy constraint. Note that in general, \ac{MB} distributions perform well on the AWGN channel, see, e.g.,
\cite[Table I]{bocherer_bandwidth_2015}.
\end{rem}

\begin{figure}[htb]
\footnotesize
\centering
\includegraphics{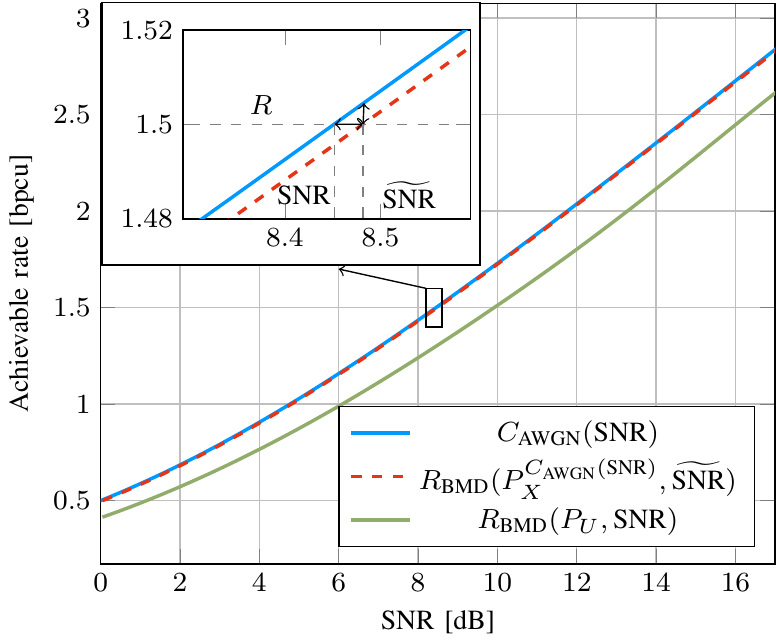}
\caption{Comparison of $R_\BMD$ to $C_\tawgn(\SNR)$ for uniform and shaped input distributions on 16-ASK.}
\label{fig:power_rate}
\end{figure}

Fig.~\ref{fig:power_rate} depicts the achievable \ac{BMD} rate performance for the 
considered scenario. The points $(\widetilde{\SNR},R)$ on the dashed red curve are given by:
\begin{enumerate}
 \item Calculate $P_X^R$ such that $R = \entr(X) - \frac{3}{4} \stackrel{!}{=} C_\tawgn(\SNR)$.
 \item Calculate $\widetilde{\SNR}$ with $\widetilde{\SNR} = R_\tbmd^{-1}(P_X^R,R)$. 
\end{enumerate}
For reference, we also plot the uniform \ac{BMD} rate, which is denoted as $R_\tbmd(P_U,\SNR)$ and $P_U(x) = 1/16, \forall x\in\cX$.

\begin{figure}[htb]
\footnotesize
\centering
\includegraphics{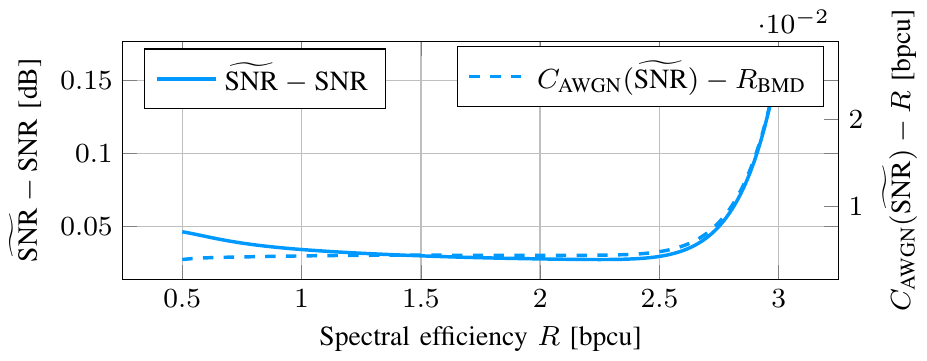}
\caption{Comparison of the $\SNR$ and rate gaps for the considered range of \SIrange{0.7}{2.7}{\bpcu}.}
\label{fig:gaps}
\end{figure}

We observe that the loss due to \ac{BMD} is negligible. This is also verified in Fig.~\ref{fig:gaps}, which depicts the 
$\SNR$ and rate gaps.

\subsection{Protograph-Based LDPC Code Ensembles}
\ac{LDPC} codes are linear block codes with a sparse $(n-k)\times n$ parity-check matrix $\vH$. The matrix 
$\vH$ can be represented by a Tanner graph~\cite[Sec. 3.3]{richardson2008modern}
consisting of variable nodes $V_i, i\in\{1,\ldots,n\}$ and check nodes $C_j, j\in\{1,\ldots,n-k\}$. A variable 
or check node has degree $d$ if it is connected to $d$ check or variable nodes, respectively. \ac{LDPC} code 
ensembles are usually characterized by the degree profiles of the variable and check nodes. For instance, 
$\lambda(x)=\sum_{d=1}^{d_v} \lambda_d x^{d-1}$ and $\rho(x)=\sum_{d=1}^{d_c} \rho_d x^{d-1}$ are the 
edge-perspective variable and check node degree polynomials with maximum degree $d_v$ and $d_c$, respectively. 
However, the degree profiles can not characterize the mapping of variable nodes to the $m$ different bit-channels
resulting from our adapted BICM transmission scheme, see Fig.~\ref{fig:bmd_channel_model}. In the following, we use the
structured ensembles of protographs~\cite{thorpe_protograph} to incorporate the bit-mapping in our threshold analysis. 
Protographs can be seen as a special instance of multi-edge type \ac{LDPC} codes~\cite{richardson_MET}, where each edge 
represents an individual edge type.

Parity-check matrices are constructed from protographs as follows. Starting from a small bipartite graph 
represented via its basematrix $\vA=[a_{lk}]$ of size $M\times N$, where $a_{lk}$ represents the number 
of edges between the protograph variable node $V_k, k\in\{1,\ldots,N\}$ and the protograph check node 
$C_l, l\in\{1,\ldots,M\}$, one applies a copy-and-permute operation (also known as lifting) to create 
$Q$ instances of the small graph and then permutes the edges so that the local edge connectivity remains
the same. The $Q$ replicas of variable node $V_k$ must be connected only to replicas of the neighbors of 
$V_k$ while maintaining the original degrees for that specific edge. The resulting bipartite graph representing 
the final parity-check matrix $\vH$ possesses $n = Q \cdot N$ variable nodes and $n-k = Q\cdot M$ check nodes. 
Parallel edges are allowed, but must be resolved during the copy-and-permute procedure.

\subsection{Code Design for Surrogate Channels}

As pointed out in our previous work \cite[Sect.~IV]{steiner_protographbased_2016}, the usual 
design approaches for \ac{LDPC} codes fail for the \ac{BICM} channel because they rely on
the assumption of binary-input symmetric-output channels so that the transmission of the all-zero 
codeword can be assumed for the analysis. The \ac{BICM} channel densities $p_{L_i|B_i}$ do not exhibit this
property, in particular when shaping is taken into account. Instead, \cite{steiner_protographbased_2016} 
pursues an approach based on \emph{surrrogate channels}, i.e., each \ac{BICM} channel $p_{L_i|B_i}$ is replaced
by a binary-input symmetric-output \ac{AWGN} channel $\tilde L_i = \tilde B_i + \tilde N_i$ with $\tilde B_i$ 
uniform on $\{\pm \frac{2}{\sigma^2_i}\}$ and $\tilde N_i$ zero mean Gaussian with 
variance $\frac{4}{\sigma_i^2}$. The parameter $\sigma_i^2$ is now chosen such that
\begin{equation}
 \entr(\tilde B_i|\tilde L_i) \stackrel{!}{=} \entr(B_i|L_i), \quad i = 1, 2, \ldots, m\label{eq:uncertainty_match}.
\end{equation}
Using the surrogates, density evolution or methods based on \ac{EXIT} charts can 
readily be applied. For our protograph based design, we resort to the PEXIT 
approach~\cite{liva_protograph_2007} to determine the asymptotic decoding threshold 
of a protograph ensemble. In the following, the notion of the 
asymptotic decoding threshold for $m$ parallel binary input AWGN channels is made precise. 

We allow different variable node degrees for each of the $m$ 
distinct bit-channels. In order to have up to $D$ different degrees 
per bit-channel, the protograph matrix $\vA$ must have at least $N=D\cdot m$ 
variable nodes. We introduce a mapping function of the form
$T(k) = \left\lceil k/D\right\rceil$ to relate each protograph variable node $V_k$ to 
a corresponding bit-level $T(k) \in\{1,\ldots,m\}$. Let $\vsigma_{\SNR,P_X^R} = \vect{\sigma_{T(1)},\ldots,\sigma_{T(k)}}^\mathrm{T}$
be the set of surrogate channel parameters for a specific $\SNR$ and \ac{SE} 
$R$ and let $I_{\text{app},k}^\ell$ be the a-posteriori mutual information of the 
$k$-th variable node in the $\ell$-th PEXIT iteration. The $\SNR$ convergence 
region $\cD(\vA,P_X^R)$ of the protograph $\vA$ and using $P_X^R$ is then given as
\[
 \cD(\vA,P_X^R) = \left\{\SNR\left|\vsigma_{\SNR,P_X^R}: I_{\text{app},k}^\ell \xrightarrow{\ell\to\infty} 1, \forall k \right.\right\}.
\]
The asymptotic decoding threshold follows as
\begin{equation}
 \SNR^\text{th}(\vA,P_X^R) = \min \cD(\vA,P_X^R)\label{eq:snr_th}.
\end{equation}

\subsection{Differential Evolution and Optimization Metric}
\label{sec:differential_evo}

In order to find the protograph ensemble with the best decoding threshold, we employ 
differential evolution~\cite{storn_differential_1997}. As the entries $a_{lk}$ of the protograph 
matrix $\vA$ are integers, the original formulation has to be adopted as shown in 
\cite{pradhan_deterministic_2013, uchikawa_design_2014}. The asymptotic decoding threshold~\eqref{eq:snr_th}
is used as a metric for the selection of new population members, if the code is optimized for a single \ac{SE}:
A member $\vA^{(g-1)}$ of the previous generation $(g-1)$ is replaced by its potential successor 
$\tilde\vA^{(g)}$ if and only if $\tilde\vA^{(g)}$ has a smaller asymptotic decoding threshold than $\vA^{(g-1)}$. 
For the robust design, we advocate for the metric
\begin{align}
 \vA^{(g)} = \argmin_{\vA\in\{\vA^{(g-1)}, \tilde\vA^{(g)}\}} \max_{R\in\cP} \SNR^\tth(\vA,P_X^R)-R^{-1}_\tbmd(P_X^R,R)\label{eq:optim_metric},
\end{align}
where $\cP\subseteq\left[\num{0.7};\num{2.7}\right]$ is the set of all considered operating points for which the code should be optimized.

\section{Simulation Results}
\label{sec:sim_results}

We design a rate 13/16 code which allows up to $D=4$ different variable node degrees 
per bit-level such that the resulting base matrices have dimensions $3\times 16$. 
The number of parallel edges is limited to 3, which results in a maximum variable 
node degree of 9. The number of degree 2 variable nodes is limited to one column
and all other nodes must have a degree of at least 3 to ensure a linear growth
of the minimum distance~\cite{nguyen_design_2012}.
We optimize basematrices for five scenarios. The first four target specific
\acp{SE} of \SI{0.7}{\bpcu}, \SI{1.1}{\bpcu}, \SI{2.1}{bpcu} and \SI{2.7}{\bpcu}, 
respectively. The last protograph $\vA_\text{rob}$ represents the robust approach that targets 
all rates in the interval $\left[0.7;2.7\right]$ jointly and is given as
\begin{align*}
\vect{3 &1 &1 &2 &1 &2 &2 &1 &0 &1 &1 &1 &3 &1 &1 &1\\ 3 &2 &2 &0 &2 &2 &2 &2 &2 &0 &1 &1 &3 &2 &1 &2\\ \undermat{B_2}{3 &0 &0 &1} &\undermat{B_3}{0 &0 &0 &0} & \undermat{B_4}{1 & 2 & 3 & 3} &\undermat{B_1}{3 & 0 & 0 & 0}}
\end{align*}
\vspace{\baselineskip}

For its optimization, we chose $\cP = \left\{0.7,1.1,2.1,2.7\right\}$. Including further rates did not improve 
the asymptotic decoding thresholds considerably -- using less or other operating points resulted in 
inferior performance. We observed good results by including the boundaries of the operating region and 
pursuing the following heuristic approach: For each \ac{SE} $R$ consider the set of surrogate channel parameters 
$\vsigma_{\SNR,P_X^R}$, which exhibit a certain ordering (``quality'') of the individual $m$ bit channels. If this ordering 
changes at a certain \ac{SE} $R$ compared to previous ones, $R$ should be added to $\cP$. In the
considered example, the ordering of $\sigma_1$ and $\sigma_4$ changes around $R\approx\SI{1.1}{\bpcu}$ and 
again $\sigma_1$ and $\sigma_3$ at around $R\approx\SI{2.1}{\bpcu}$. For small protograph sizes and limited degrees
of freedom, it may suffice to consider the boundary operating points only.

The protographs $\vA_{R}, R\in\{0.7,1.1,2.1,2.7\}$ as well as their parity-check-matrices
can be found online at~\cite{website_istc2016}.

\begin{figure}[htb]
\footnotesize
\centering
\includegraphics{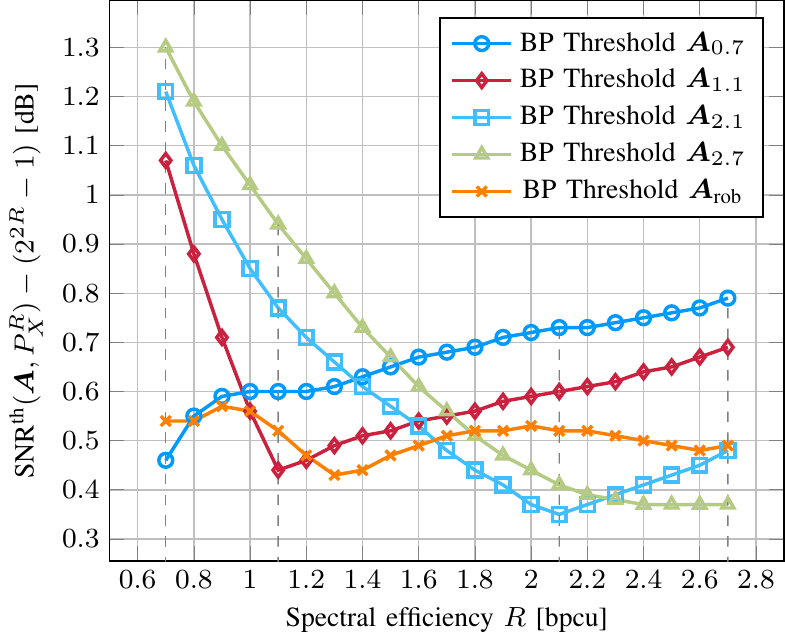}
\caption{Gap in \si{dB} to $C_\tawgn(\SNR)$ of the asymptotic protograph \ac{BP} decoding thresholds \eqref{eq:snr_th} over the range of considered spectral efficiencies.}
\label{fig:fer_gap_curves_asymptotic}
\end{figure}

Fig.~\ref{fig:fer_gap_curves_asymptotic} depicts the $\SNR$ gap of the asymptotic decoding thresholds to \ac{AWGN}
capacity, i.e., $\SNR^\tth(\vA,P_X^R)-(2^{2R}-1)$, in \si{dB} for the considered range of \acp{SE}. 
Protographs which are optimized for one particular \ac{SE} tend to perform poor if operated at other \acp{SE}.
This is particularly the case for the codes optimized for high \acp{SE}, where the gap increases up to \SI{1.3}{dB}
when operated at lower $\SNR$. The robust protograph design exhibits the desired feature of minimizing the maximum gap for each operating point
in $\cP$ and therefore achieves a balanced behavior.

\begin{figure}[htb]
\footnotesize
\centering
\includegraphics{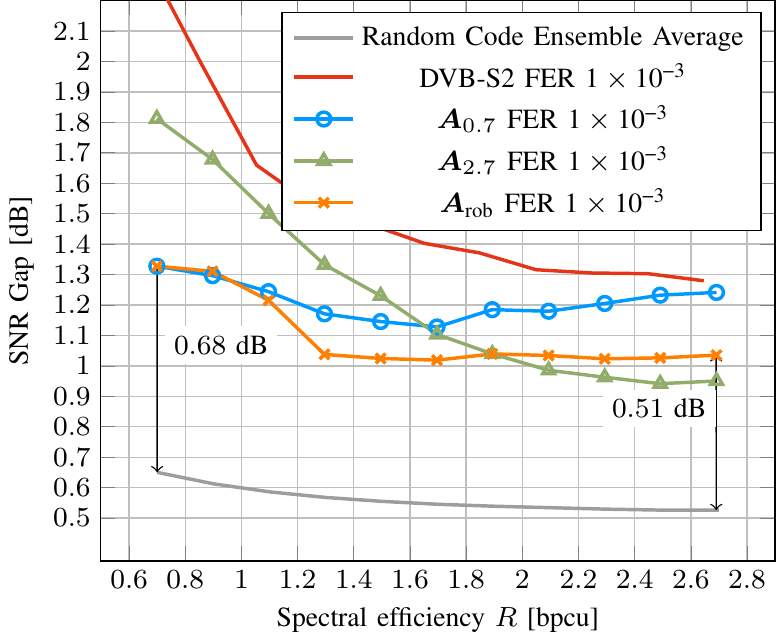}
\caption{Gap in \si{dB} to $C_\tawgn(\SNR)$ for different protograph designs over the range of considered spectral efficiencies at a target \ac{FER} of \num{e-3}.}
\label{fig:fer_gap_curves}
\end{figure}

For a finite length comparison in Fig.~\ref{fig:fer_gap_curves}, the protographs have first been lifted by a factor of three to 
remove parallel edges and then by a factor of \num{338} to yield parity-check matrices of size $2705\times 16224$.
As a baseline for performance comparison, we choose the 5/6 DVB-S2 code for short frame sizes, 
which has a blocklength of $n = 16200$~\cite{etsi2009dvb}. For the bit-mapping, bit-levels two to 
four are assigned consecutively to the first 12150 variable nodes, whereas bit-level one is assigned 
to the remaining ones. As the information part of the parity-check matrix has mostly degree three variable nodes
and only a small number of 360 degree 13 variable nodes, optimizing the bit-mapper did not improve
performance.
In addition to the DVB-S2 reference, we also plot the random coding bound $2^{-n_\tc E_G(R)}$~\cite{gallager1968},
which provides an upper bound on the frame error performance of a random code ensemble for a given blocklength.
Hereby, $E_G(R)$ denotes the Gallager exponent for rate $R$.

In all cases, $100$ \ac{BP} iterations with a full sum-product update rule are performed. We observe that the predictions of the asymptotic decoding thresholds are well reflected
in the finite-length performance as well.

\section{Conclusion}
\label{sec:conclusion}

In this work, we proposed a novel, robust design approach for protograph based \ac{LDPC} codes for rate-adaptive communication via probabilistic
amplitude shaping. Using one 16-ASK constellation and one 13/16 rate code, any \ac{SE} between \SIrange{0.7}{2.7}{\bpcu} can be achieved with a gap of at most
\SI{1.3}{dB} for a target \ac{FER} of \num{e-3} and a blocklength of \num{16224}. Future research should focus on investigating other robust optimization metrics, 
such as the rate-backoff at the decoding threshold and the influence of the choice of operating points in the set $\cP$.


\end{document}